\documentclass{iau}
\usepackage{graphicx}
\usepackage{natbib}

\title[Magnetic fields in stars with radiative envelopes] 
{Modeling surface magnetic fields in stars with radiative envelopes}

\author[Oleg Kochukhov]   
{Oleg Kochukhov}

\affiliation{Department of Physics and Astronomy, Uppsala University \\ Box
515, SE-75120 Uppsala, Sweden \\email: {\tt oleg.kochukhov@physics.uu.se}}

\pubyear{2013}
\volume{302}  
\pagerange{1--10}
\jname{Magnetic Fields throughout Stellar Evolution}
\editors{A.C. Editor, B.D. Editor \& C.E. Editor, eds.}
\begin{document}

\maketitle

\begin{abstract}
Stars with radiative envelopes, specifically the upper main sequence chemically peculiar (Ap) stars, were among the first objects outside our solar system for which surface magnetic fields have been detected. Currently magnetic Ap stars remains the only class of stars for which high-resolution measurements of both linear and circular polarization in individual spectral lines are feasible. Consequently, these stars provide unique opportunities to study the physics of polarized radiative transfer in stellar atmospheres, to analyze in detail stellar magnetic field topologies and their relation to starspots, and to test different methodologies of stellar magnetic field mapping. Here I present an overview of different approaches to modeling the surface fields in magnetic A- and B-type stars. In particular, I summarize the ongoing efforts to interpret high-resolution full Stokes vector spectra of these stars using magnetic Doppler imaging. These studies reveal an unexpected complexity of the magnetic field geometries in some Ap stars.
\keywords{stars: magnetic fields, stars: early-type, stars: chemically peculiar, polarization}
\end{abstract}

\firstsection
\section{Magnetic stars with radiative envelopes}

Observational manifestations of magnetic fields in intermediate- and high-mass stars with radiative envelopes differ considerably from the magnetism of solar-type and low-mass stars. As directly observed for the Sun and inferred for many late-type stars, vigorous envelope convection and differential rotation give rise to ubiquitous intermittent magnetic fields, which evolve on relatively short time-scales and generally exhibit complex surface topologies. Although details of the dynamo operation in late-type stars, in particular the relative importance of the convective and tachocline dynamo mechanisms is a matter of debate \citep{brandenburg:2005} and probably depends on the position in the H-R diagram, it is understood that essentially every cool star is magnetic. Chromospheric and X-ray emission and surface temperature inhomogeneities, which are responsible for characteristic photometric variability, provide an indirect evidence of the surface magnetic fields in cool stars.

In contrast, stars hotter than about mid-F spectral type and more massive than $\sim$\,1.5$M_\odot$ are believed to lack a sizable convective zone near the surface\footnote{Very massive stars may generate magnetic fields in the sub-surface Fe convection zone \citep{cantiello:2011}. However, no signatures of these fields have been discovered so far \citep{kochukhov:2013b}.} and therefore are incapable of generating observable magnetic fields through a dynamo mechanism. Nevertheless, about 10\% of O, B, and A stars exhibit very strong (up to 30~kG), globally organized (axisymmetric and mostly dipolar-like) magnetic fields that appear to show no intrinsic temporal variability whatsoever. This phenomenon is usually attributed to the so-called fossil stellar magnetism -- a hitherto unknown process (possibly related to initial conditions of stellar formation or early stellar mergers) -- by which a fraction of early-type stars become magnetic early in their evolutionary history. By far the most numerous among the early-type magnetic stars are the A and B magnetic chemically peculiar (Ap/Bp) stars. These stars were the first objects outside our solar system in which the presence of magnetic field was discovered \citep{babcock:1947}. Ap/Bp stars are distinguished by slow rotation \citep{abt:1995} and are easy to recognize spectroscopically by the abnormal line strengths of heavy elements in their absorption spectra. These spectral peculiarities are related to distinctly non-solar surface chemical composition of these stars and non-uniform horizontal \citep[e.g.][]{kochukhov:2004e,nesvacil:2012} and vertical distributions of chemical elements \citep[e.g.][]{ryabchikova:2002,kochukhov:2006b}. These chemical structures are presumably formed by the magnetically-controlled atomic diffusion \citep{alecian:2010} operating in stable atmospheres of these stars.

The chemical spot distributions and magnetic field topologies of Ap stars remain constant (frozen in the atmosphere). Yet, all these stars show a pronounced and strictly periodic (with periods from 0.5~d to many decades) spectroscopic, magnetic and photometric variability due to rotational modulation of the aspect angle at which stellar surface is seen by a distant observer. A subset of cool magnetic Ap stars -- rapidly oscillating Ap (roAp) stars -- also varies on much shorter time scales ($\sim$10~min) due to the presence of $p$-mode oscillations aligned with the magnetic field \citep{kurtz:2000}.

A large field strength and lack of intrinsic variability facilitates detailed studies of the field topologies of individual magnetic Ap stars and statistical analyses of large stellar samples. In this review I outline common methodologies applied to detecting and modeling surface magnetic fields in early-type stars and summarize main observational results. Closely related contributions to this volume include an overview of massive-star magnetism (Wade, Grunhut), a discussion of the stability and interior structure of fossil magnetic fields (Braithwaite), and an assessment of the chemical peculiarities and magnetism of pre-main sequence A and B stars (Folsom).

\section{Magnetic field observables for early-type stars}

With a few exceptions, investigations of the magnetism of cool stars have to rely on high-resolution spectropolarimetry and to engage in a non-trivial interpretation of the complex polarization signatures inside spectral lines in order to characterize the field topologies \citep{donati:1997}. In contrast, a key advantage of the magnetic field studies of early-type stars with stable global fields is availability of a wide selection of magnetic observables that are simple to derive and interpret, but are still suitable for a coarse analysis of the surface magnetic field structure.

The simplest approach to detecting the presence of the field in early-type stars is to perform spectroscopic observation with a Zeeman analyzer equipped with a quarter-wave retarder plate and a beamsplitter. The resulting pair of left- and right-hand circularly polarized spectra will exhibit a shift proportional to the Land\'e factors of individual spectral lines and to \textit{the mean longitudinal magnetic field} -- the line-of-sight field component averaged over the stellar disk. Various versions of this longitudinal field diagnostic technique have been applied by \citet{babcock:1958}, \citet{kudryavtsev:2006}, and \citet{monin:2012} to medium-resolution spectra. \citet{landstreet:1980} and \citet{bagnulo:2002a} have extended it to, respectively, photopolarimetric and low-resolution spectropolarimetric measurements of polarization in the wings of hydrogen lines.

The mean longitudinal magnetic field represents a particular example of an integral measurement derived from a moment of Stokes $V$ profile (the first moment in this case). \citet{mathys:1989} have generalized the moment technique to other Stokes $I$ and $V$ profile moments. In practice, only \textit{the mean quadratic field} (the second moment of Stokes $I$) and \textit{crossover} (the second moment of Stokes $V$), in addition to longitudinal field, were systematically studied by Mathys and collaborators using medium-resolution observations of Ap stars \citep[e.g.][]{mathys:1997a}.

\begin{figure}[t]
\begin{center}
\includegraphics[width=\textwidth]{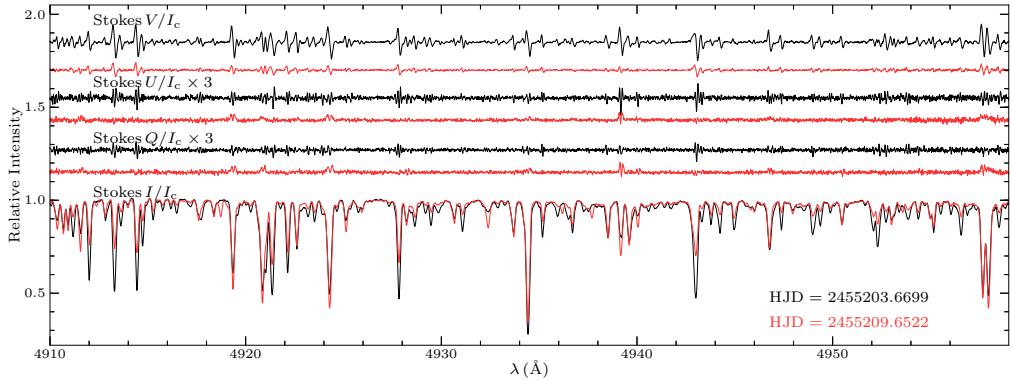} 
\caption{Representative four Stokes parameter Ap-star observations obtained with the HARPSpol spectropolarimeter at the 3.6-m ESO telescope. This figure shows a small segment of the echelle spectra of the cool Ap star HD\,24712 at two different rotational phases. The Stokes $QUV$ observations are offset vertically. Clear circular and linear polarization signatures are evident in many individual spectral lines. Detailed analysis of these observations has been published by \citet{rusomarov:2013}.}
\label{fig1}
\end{center}
\end{figure}

The three aforementioned magnetic observables can be related to the disk-averaged properties of stellar magnetic field under a number of simplifying and restrictive assumptions (weak lines, weak field, no chemical spots). At the same time, an entirely assumption-free method to diagnose magnetic fields in early-type stars is to measure a separation of the resolved Zeeman-split components in the intensity profiles of magnetically sensitive spectral lines. The resulting \textit{mean field modulus} measurements have been obtained for many Ap stars showing strong magnetic fields in combination with a particularly slow rotation \citep{mathys:1997b}.

A different approach can be applied to diagnose the transverse magnetic field components which give rise to linear polarization. Cooler Ap stars exhibit measurable \textit{net linear polarization} due to differential saturation of the $\pi$ and $\sigma$ components of the strong spectral lines. The resulting net $Q$ and $U$ signals can be detected with a broad-band photopolarimetric technique and related to disk-averaged characteristics of the transverse magnetic field \citep{landolfi:1993}.

With the advent of high-resolution spectropolarimeters at the 2--4~m-class telescopes it became possible to directly record and interpret the circular and linear polarization signatures in individual spectral lines. A multi-line LSD (least-squares deconvolution) technique \citep{donati:1997,kochukhov:2010a} is often used in conjunction with such observations to obtain very high signal-to-noise (S/N) ratio mean intensity and polarization profiles. LSD analysis greatly facilitates detection of weak magnetic fields \citep[e.g.][]{auriere:2007} and allows to derive the mean longitudinal field, net linear polarization and other profile moments for direct comparison with historic studies. The first high-resolution full Stokes vector investigations were carried out for Ap stars with now decommissioned MuSiCoS spectropolarimeter \citep{wade:2000b,bagnulo:2001}. More recent analyses took advantage of ESPaDOnS, NARVAL \citep{silvester:2012} and HARPSpol \citep{rusomarov:2013} instruments. An example of exceptionally high quality ($R>10^5$, $S/N>400$, 16 rotation phases, coverage of 3800--6910~\AA\ wavelength region; see \citealt{rusomarov:2013} for details) HARPSpol Stokes spectra of the roAp star HD\,24712 is illustrated in Fig.~\ref{fig1}. These observations represent the highest quality full Stokes vector spectra available for any star other than the Sun while covering a much wider wavelength domain than typical for solar polarization observations.

\section{Multipolar modeling of magnetic observables}

Fitting the phase curves of one or several magnetic observables constitutes the basic method of constraining the stellar magnetic field parameters. A limited information content of integral observables and their relatively simple sinusoidal variation in most Ap stars justifies describing  the stellar magnetic field topology with a small number of free parameters. By far the most common approximation is a simple rigidly rotating dipolar field geometry \citep{stibbs:1950}, characterized by an inclination angle of the stellar rotational axis $i$, magnetic obliquity $\beta$, and a polar field strength $B_{\rm p}$. Observations of the phase variation of the mean longitudinal magnetic field alone allows one to constrain $B_{\rm p}$ and $\beta$, provided that $i$ is known and not too close to 90$^{\rm o}$. In the latter case longitudinal field measurements constrain only the product $B_{\rm p} \sin\beta$.

Occasional deviations of the longitudinal field curves from the sinusoidal shape expected for a dipolar field and the requirement to fit simultaneous measurements of the longitudinal and mean surface fields led to the development of more complex field geometry models, described with additional free parameters. Different low-order multipolar field parameterizations have been considered in the literature. This included a dipolar field offset along its axis \citep[e.g.][]{preston:1971a}, an arbitrary offset dipole \citep{townsend:2005}, an axisymmetric combination of the aligned dipole, quadrupole, and octupole components \citep{landstreet:1988}, a general non-axisymmetric quadrupolar field \citep{bagnulo:1996}, and a potential field geometry formed by a superposition of an arbitrary number of point-like magnetic sources \citep{gerth:1999}. 

The choice between these different multipolar parameterizations is typically subjective. Stellar observations themselves frequently do not allow one to make a clear-cut distinction between multipolar models established in the framework of different parameterizations, even when several magnetic observables are available for a given star. Indeed, it was demonstrated that the same set of observed magnetic curves can be successfully interpreted with very different actual surface magnetic field distributions, depending on which multipolar parameterization is used \citep{kochukhov:2006c}.

Nevertheless, systematic applications of multipolar modeling to a large number of stars allowed to reach interesting conclusions. Using  centered dipole fits to the longitudinal field curves, \citet{auriere:2007} established the existence of a lower field limit of $B_{\rm p}\approx300$~G for Ap stars. This threshold of global fossil field strength is likely to be of fundamental importance for understanding the magnetism of intermediate-mass stars (see Ligni{\`e}res, this volume). Among other notable findings one can mention the work by \citet{landstreet:2000}, who demonstrated that magnetic field axis tends to be more aligned with the stellar rotation axis for Ap stars with long ($>25$~d) rotation periods. \citet{bagnulo:2002} confirmed this result using a different multipolar field parameterization. They also found a certain dependence of the relative orientation of the dipolar and quadrupolar components on the stellar rotation rate. Both studies fitted the observed curves of the mean field modulus, longitudinal field, crossover, and quadratic field. Despite the overall statistical agreement, in many individual cases the surface field maps resulting from the application of Landstreet's and Bagnulo's parameterizations appear very different for the same stars. Furthermore, some of the observables are poorly reproduced by either multipolar model, which can be ascribed to the presence of more complex field structures, an unaccounted influence of chemical abundance spots or to shortcomings of the basic assumptions of the moment technique or to a combination of all these effects.

Some applications of the multipolar fitting procedure have incorporated detailed polarized radiative transfer (PRT) modeling of the Zeeman-split Stokes $I$ profiles into solving for the surface field geometry \citep{landstreet:1988,bailey:2011}. This approach enables an independent validation of the magnetic field topology and makes possible to deduce a schematic horizontal distribution of chemical spots in addition to studying the field geometry. However, a feedback of chemical spots on the magnetic observables is not taken into account by these studies.

The most sophisticated and well-constrained non-axisymmetric multipolar models were developed by \citet{bagnulo:2000,bagnulo:2001} for the Ap stars $\beta$~CrB and 53~Cam using all integral magnetic observables available from the Stokes $IV$ spectra together with the broad-band linear polarization measurements. However, even such detailed models do not guarantee a satisfactory description of the same Stokes parameter spectra from which the magnetic observables are obtained. As found by \citet{bagnulo:2001}, the multipolar models derived from magnetic observables provide a rough qualitative reproduction of the phase variation of the Stokes $V$ profiles but sometimes fail entirely in matching the Stokes $QU$ signatures observed in individual metal lines. This problem points to a significant limitation of the multipolar models: a successful fit of the phase curves of all magnetic observables is often non-unique and is generally insufficient to guarantee an adequate description of the high-resolution polarization spectra. On the other hand, discrepancies between the model predictions and observations in partially successful multipolar fits cannot be easily quantified in terms of deviations from the best-fit magnetic field geometry model.

\section{Interpretation of Stokes parameter spectra}

Modeling of high-resolution observations of polarization signatures in individual spectral lines or in mean line profiles represents the ultimate method of extracting information about stellar magnetic field topologies. The wide-spread usage of the LSD processing of high-resolution polarization spectra stimulated development of various Stokes $V$ profile fitting methodologies \citep{alecian:2008a,grunhut:2012,petit:2012}. These studies usually deal with weak-field early-type stars without prominent chemical spots (e.g. magnetic massive or Herbig Ae/Be stars, but not typical Ap stars). The observed LSD profiles are approximated with a dipolar field topology, using a simplified analytical treatment of the polarized line formation. Eventual variations caused by chemical spots or other surface features are not considered. So far, this modeling approach has been applied to a few stars, but it has a potential of providing constraints on magnetic field and other stellar parameters (inclination, $v\sin i$) beyond what can be obtained from the longitudinal field curves \citep{shultz:2012,kochukhov:2013a}.

A more rigorous approach to the problem of finding the stellar surface magnetic field geometry from spectropolarimetric observations is to perform a full magnetic inversion known as Magnetic (Zeeman) Doppler imaging (MDI). In the MDI methodology developed by \citet{piskunov:2002a} and \citet{kochukhov:2002c} the time-series observations in two or four Stokes parameters are interpreted with detailed PRT calculations, taking surface chemical inhomogeneities into account. Simultaneous reconstruction of the magnetic field topology and chemical spot distributions is carried out by solving a regularized inverse problem. Regularization limits the range of possible solutions and is needed to stabilize the iterative optimization process and to exclude small-scale surface structures not justified by the data. Different versions of regularization have been applied for magnetic mapping of early-type stars. \citet{piskunov:2002a} needed only the local Tikhonov regularization (imposing a correlation between neighboring surface pixels) to achieve a reliable reconstruction of an arbitrary magnetic field map from full Stokes vector spectra. However, a more restrictive multipolar regularization \citep{piskunov:2002a} or a spherical harmonic field expansion \citep{donati:2006b} is required to reconstruct a low-order multipolar field in the case when only Stokes $I$ and $V$ observations are available.

Magnetic imaging of Ap stars was recently coupled to a calculation of the atmospheric models that take into account horizontal variations of the atmospheric structure due to chemical spots \citep{kochukhov:2012}. However, while the self-consistency between spots and atmospheric models is critical for magnetic mapping of cool stars \citep{rosen:2012} and may be needed for mapping He inhomogeneities in He-rich stars, it is generally unnecessary for treating metal spots in Ap stars.

It should be emphasized that, in contrast to temperature or chemical spot imaging from intensity spectra, the MDI with polarization data is not limited to rapid rotators. Polarization is strongly modulated by the stellar rotation even for magnetic stars with negligible $v\sin i$. Numerical experiments and studies of real stars demonstrated that this modulation is sufficient for recovering the field structure at least at the largest spatial scales \citep[e.g.][]{kochukhov:2002c,donati:2006b}.

Several studies \citep{kochukhov:2002b,luftinger:2010,rivinius:2013} applied MDI to high-quality time-series circular polarization spectra of several early-type magnetic stars. These Stokes $V$ analyses did not find any major deviations from dipolar field topologies. At the same time, they found numerous examples of chemical spot maps showing diverse and complex distributions of chemical elements, often not correlating in any meaningful way to the underlying simple magnetic field geometry. These results are difficult to explain in the framework of atomic diffusion theory because the latter expects a very similar behavior for different elements and a definite correlation between the spots and magnetic field \citep{alecian:2010}.

A couple of other studies have attempted to examine the surface magnetic field structure in B-type stars with fields deviating significantly from dipolar geometry. A study of the He-peculiar star HD\,37776 \citep{kochukhov:2011a} has simultaneously interpreted a longitudinal field curve and moderate-resolution Stokes $V$ spectra. This analysis inferred a decisively non-axisymmetric, complex and strong (up to 30 kG locally) magnetic field, but ruled out a record $\sim$\,100~kG quadrupolar field proposed for this star by previous longitudinal field curve fits. An MDI study of the early B-type star $\tau$~Sco \citep{donati:2006b} revealed the presence of weak complex magnetic field configuration, which exhibits no appreciable temporal variation \citep{donati:2009}. These two studies have proven that stable complex fields can exist in early-type stars and tend to be found in the most massive objects. Despite these impressive MDI results, it should be kept in mind that the Stokes $IV$ inversions are intrinsically non-unique and their outcome is highly sensitive to additional constraints adopted to stabilize inversions. Details of the magnetic field maps of HD\,37776 and $\tau$~Sco are likely to change if different regularizations or different forms of spherical harmonic expansion are adopted for magnetic imaging.

Numerical tests of MDI inversions \citep{donati:1997,kochukhov:2002c} have concluded that reconstruction of stellar magnetic field topologies from the full Stokes vector data should be considerably more reliable and resistant to cross-talk and non-uniqueness problems in comparison to the Stokes $IV$ imaging. In particular, a four Stokes parameter inversion is able to recover the field structure without imposing any \textit{a priori} constraints on the global field geometry. The first Stokes $IQUV$ MDI studies exploiting this possibility were carried out for the Ap stars 53~Cam \citep{kochukhov:2004d} and $\alpha^2$~CVn \citep{kochukhov:2010} using the MuSiCoS spectra collected by \citet{wade:2000b}. Both studies succeeded in reproducing the phase variation of the circular and linear polarization signatures in metal lines with the magnetic maps containing small-scale deviations from the dominant dipolar-like field component. Interestingly, it was the inclusion of Stokes $QU$ profiles in the magnetic inversions that allowed to ascertain the presence of complex fields. The deviations from dipolar field configurations occur on much smaller spatial scales than can be described by a quadrupolar field. Thus, the widely adopted dipole+quadrupole expansion may not be particularly useful for interpreting the Stokes $IQUV$ spectra of Ap stars.

\begin{figure}[t]
\begin{center}
\includegraphics[width=\textwidth]{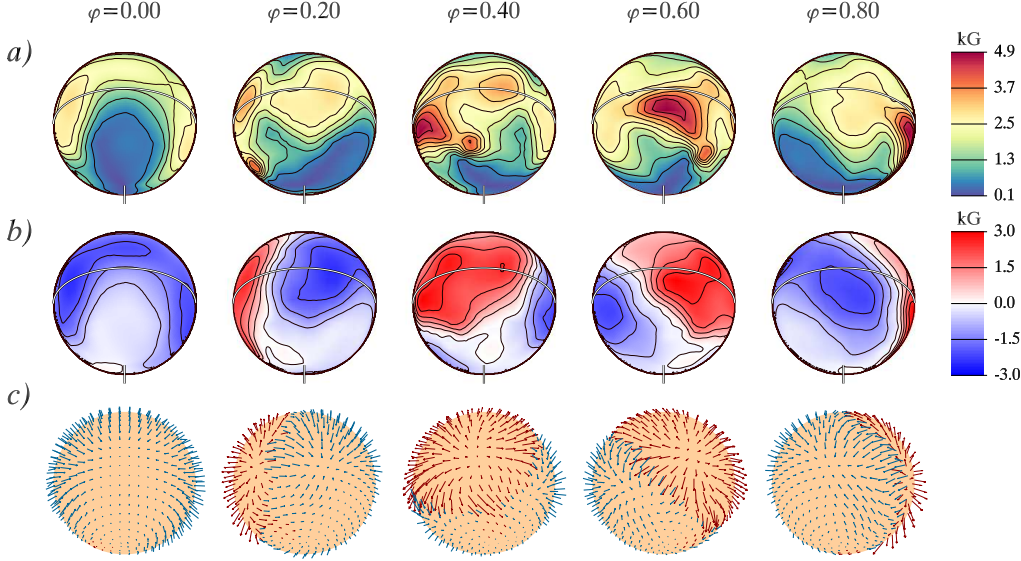} 
\caption{Magnetic field geometry for the prototypical magnetic Ap star $\alpha^2$~CVn reconstructed by Silvester et al. (2013, submitted) using MDI. These magnetic inversions were based on a set of high-resolution four Stokes parameter spectra described by \citet{silvester:2012}. The star is shown at five different rotation phases and an inclination angle $i=120^{\rm o}$. The spherical maps show a)  surface distribution of the magnetic field strength, b) distribution of the radial magnetic field component, and c) vector plot of magnetic field. The field complexity is evident, especially in the field strength map.}
\label{fig2}
\end{center}
\end{figure}

\begin{figure}[t]
\begin{center}
\includegraphics[width=\textwidth]{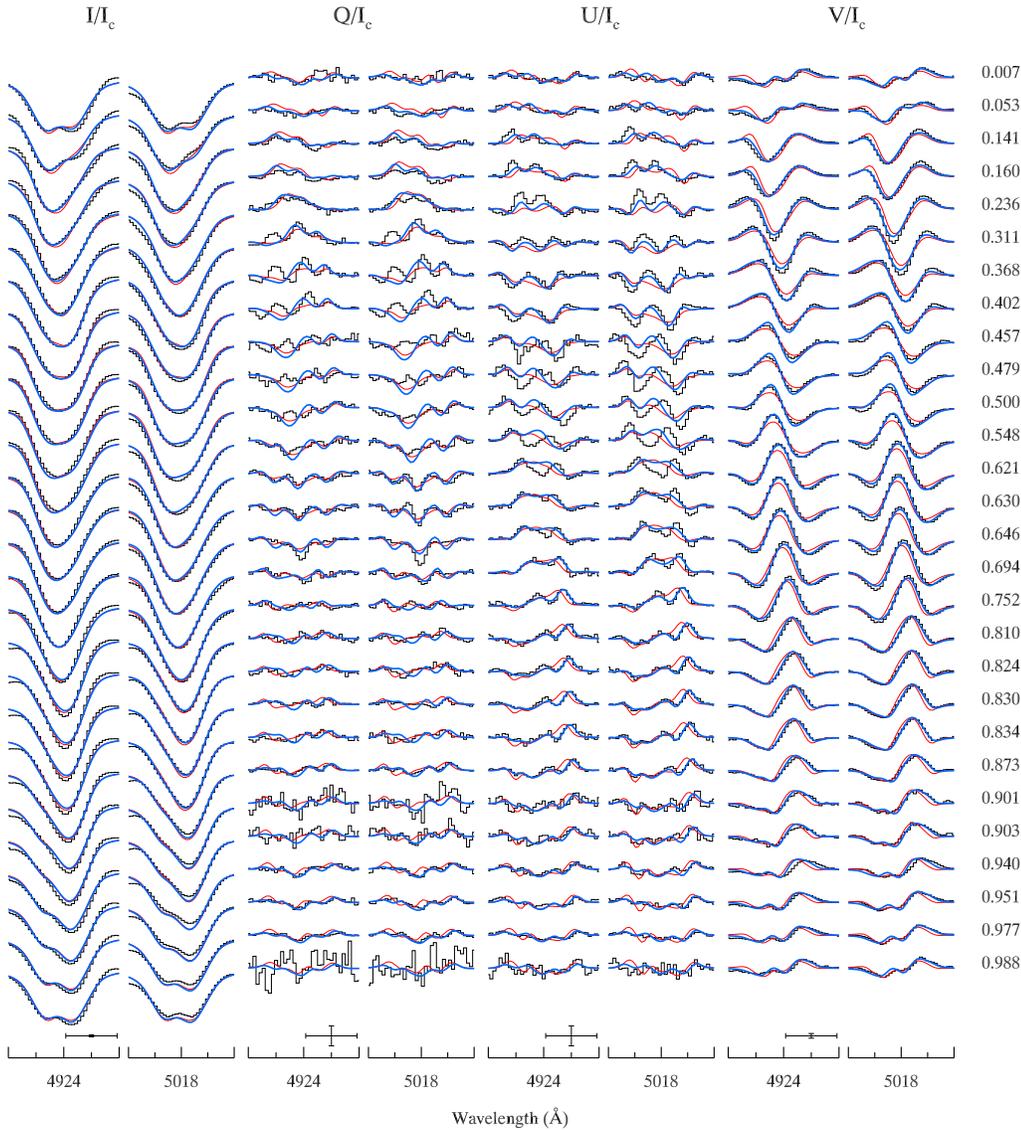} 
\caption{Attempt to reproduce the four Stokes parameter observations (histogram) of the two strong Fe~{\sc ii} lines in the spectrum of Ap star $\alpha^2$~CVn with a dipolar field model (red thin line) and a non-axisymmetric dipole plus quadrupole magnetic configuration (blue thick line). It is evident that none of the models fits the $QU$ observations at all rotational phases, thus requiring a more structured field geometry for this star (see Fig.~\ref{fig2}). However, the presence of this small-scale field cannot be recognized from the Stokes $IV$ data alone.}
\label{fig3}
\end{center}
\end{figure}

The limited resolution, S/N ratio, and wavelength coverage of the MuSiCoS spectra allowed us to model the Stokes $IQUV$ profiles of only 2--3 saturated metal lines. A new generation of MDI studies is currently underway, taking advantage of the higher-quality Stokes profile data available from ESPaDOnS, NARVAL, and HARPSpol spectropolarimeters. In particular, Silvester et al. (submitted) have reassessed the magnetic field topology of $\alpha^2$~CVn using new observations and extending the PRT MDI modeling to a large number of weak and strong Fe and Cr lines. The resulting magnetic maps (Fig.~\ref{fig2}) show some dependence of the mapping results on the spectral line choice but generally demonstrate a very good agreement with the magnetic topology found by \citet{kochukhov:2010} from observations obtained about 10 years earlier. Thus, the small-scale magnetic features discovered in Ap stars by MDI studies do not exhibit any temporal evolution. The new four Stokes parameter observations of $\alpha^2$~CVn also demonstrate very clearly the necessity of going beyond a low-order multipolar field model and the role of  Stokes $QU$ spectra in recognizing this field complexity. As illustrated by Fig.~\ref{fig3}, an attempt to reproduce the observations of $\alpha^2$~CVn with either a pure dipole or dipole+quadrupole geometries fails for Stokes $QU$ while providing a reasonable fit to Stokes $IV$.

Despite an improved sensitivity to complex fields, not all four Stokes parameter magnetic inversions point to local deviations from dipolar field topologies. The ongoing HARPSpol study of the cool Ap star HD\,24712 does not reveal any significant non-dipolar field component (\citealt{rusomarov:2013} and in this volume). The preliminary conclusion of this work is that the previous MDI analysis of this star carried out by \citet{luftinger:2010} using the Stokes $I$ and $V$ data and assuming a dipolar field did not miss any significant aspects of the field topology. HD\,24712 is much cooler, hence less massive and/or older than 53~Cam and $\alpha^2$~CVn, raising an intriguing possibility of the mass and/or age dependence of the degree of magnetic field complexity in early-type stars. The presence of very complex non-dipolar fields only in relatively massive B-type magnetic stars (HD\,37776, $\tau$~Sco) agrees with this trend.

\section{Conclusions}

Modeling of magnetic fields in early-type stars with radiative envelopes has traditionally assumed low-order multipolar field configurations and focused on interpretation of the phase curves of the mean longitudinal magnetic field and other integral magnetic observables derived from moderate-quality circular polarization spectra. The studies based on this methodology have reached several important statistical conclusions about the nature of magnetic fields in Ap and related stars. This includes the discovery of a lower threshold of the surface magnetic field strength, demonstration of an alignment of the magnetic and rotational axes in stars with long rotation periods, and confirmation of long-term stability stability of fossil magnetic fields.

As observational material improves and high-resolution spectra in several Stokes parameters become more widely available, the focus of magnetic field modeling studies gradually shifts to direct interpretation of the polarization signatures in spectral line profiles. The most powerful version of this methodology -- magnetic Doppler imaging inversions based on detailed calculation of polarized spectra -- has been applied to a handful of Ap stars observed in all four Stokes parameters. These studies revealed significant local deviations from a dominant dipolar field topology, suggesting that the magnetic field structure of early-type stars with fossil fields is more complex than thought before and that the degree of field complexity increases with stellar mass. With the exception of a couple of massive stars with distinctly non-dipolar fields, these small-scale field structures could be recognized and fully characterized only using spectropolarimetric observations in all four Stokes parameters.

\bigskip
{\it Acknowledgements.} The author is a Royal Swedish Academy of Sciences Research Fellow, supported by the grants from Knut and Alice Wallenberg Foundation and Swedish Research Council.

\end{document}